\documentclass[acmsmall,screen]{acmart}
\AtBeginDocument{%
  }

\usepackage{xcolor}
\usepackage{todonotes}
\usepackage{indentfirst}  
\usepackage{algorithm}
\usepackage{algpseudocode}
\usepackage{multirow}
\usepackage{threeparttable}
\usepackage{enumitem}

\newcommand{\toolname}{\textsc{Test2VA}}
\usepackage[dvipsnames]{xcolor}

\begin{document}

\title{Generalizing GUI Tests into Voice Assistant Capabilities through Behavioral Variation Discovery}

\author{Xue Qin}
\email{xue.qin@villanova.edu}
\orcid{0000-0003-2503-1527}
\affiliation{%
  \institution{Villanova University}
  \city{Villanova}
  \state{Pennsylvania}
  \country{USA}
}

\author{Sumesh Surendran Letha}
\email{ssuren01@villanova.edu}
\affiliation{%
  \institution{Villanova University}
  \city{Villanova}
  \state{Pennsylvania}
  \country{USA}
}

\renewcommand{\shortauthors}{Xue et al.}

\begin{abstract}
Voice Assistants (VAs) and agentic assistants increasingly support app-specific tasks through natural-language interaction.
However, these capabilities still require app-specific execution logic.
Existing approaches either derive this logic dynamically from GUI context, require developers to define app functions, or reuse GUI tests that remain tied to a single concrete task instance.
To address this limitation, we propose \toolname{}, an approach that generalizes existing GUI tests into reusable app-specific VA capabilities. \toolname{} adopts mutation as an exploration mechanism to discover plausible behavioral variations of test interactions, dynamically validates these variations on the running application with adapted assertions, and generalizes the validated task instances into capabilities that preserve the common task pattern while supporting broader user requests.
We evaluate \toolname{} on 100 Espresso GUI tests from ten real-world Android applications, covering 407 GUI events and 115 assertions. Across 396 reproduced events, \toolname{} reduces 18,999 available mutation candidates to 1,272 predicted candidates before dynamic validation, identifies mutable events with 83.33\% precision, 87.79\% recall, and an 85.50\% F1 score, and generates 65 VA capabilities. Among them, 64 execute successfully and 56 satisfy our capability quality requirements. We further analyze unsuccessful and low-quality cases to identify common causes of failure.
Finally, we conduct a user study with 18 participants and 150 task observations to evaluate the generated capabilities in real app tasks. The VA successfully completes 123 observations, achieving an overall success rate of 82.0\%, and 16 of 18 participants rate the VA as useful or very useful. These results show the feasibility of generalizing test-derived task logic into reusable app-specific VA capabilities.
\end{abstract}




\maketitle

\section{Introduction} \label{sec:intro}

Voice interaction is increasingly becoming part of a broader AI-assisted user experience.
Modern voice assistants now include traditional assistants such as Siri and Alexa as well as AI-enabled assistants such as ChatGPT Voice, Gemini Live, and Copilot Voice~\cite{report_emarketer_2026}.
Smartphones remain a primary access point for these voice-enabled AI experiences~\cite{report_emarketer_voiceai_2026}.
Mobile platforms are also enabling applications to expose functionality to AI agents and personalized assistants.
For example, Android introduced AppFunctions to allow app functionality to be invoked by AI agents~\cite{android_appfunctions_2026}.
These trends create opportunities for conversational and agentic interactions with app-specific tasks, such as recording blood pressure or adding a financial transaction.
However, such interactions require connecting natural-language requests to application-specific execution logic.
With approximately two million applications available on Google Play~\cite{appbrain_2026}, efficiently constructing these app-specific capabilities remains a significant development challenge.

Existing approaches support app-specific conversational and agentic interactions in different ways.
Research systems have explored direct voice control~\cite{vu2023voicify}, programming by demonstration~\cite{VASTA, SUGILITE}, and more recently LLM-based mobile agents that translate user requests into GUI actions~\cite{huang2025prompt2task}, reason about task execution across application pages~\cite{wu2025reachagent}, or leverage structured UI representations for app-specific automation~\cite{lin2025uicompass}.
Mobile platforms are also beginning to provide explicit interfaces for exposing app functionality to agents.
For example, Android AppFunctions allows developers to define application functions that agents can discover and invoke~\cite{android_appfunctions_2026}.
However, such explicit integrations still require developers to identify app-specific capabilities and implement their execution logic and parameters, while GUI-based agents derive this logic dynamically from runtime context.
Another line of work reuses existing GUI tests as grounded specifications of app-specific, user-facing tasks~\cite{qin2025apptova}.
However, this approach directly converts the original test method into a VA capability, so the generated capability remains tied to the concrete task instance encoded in that test.
We address this limitation by generalizing test-derived app-specific capabilities so that they can support broader task variations beyond the single concrete instance.

GUI tests encode concrete, user-facing interaction sequences that exercise app-specific functionality and verify expected outcomes. This makes them a promising source of execution knowledge for constructing app-specific VA capabilities. However, a test usually captures only one concrete task instance. 
For example, a test for a calling function may call a specific contact such as ``Bob,'' even though the underlying app task is to call any contact selected by the user.
Directly reusing the original test therefore limits the resulting VA capability to the behavior explicitly encoded in that test.
We hypothesize that a GUI test can represent more than a single concrete execution: some interactions capture the common task structure, while others reflect instance-specific choices that can vary across related task instances. If these valid variations can be identified and validated, the original test can be generalized into a broader VA capability while preserving its grounded execution logic.
Based on this hypothesis, this paper reuses existing GUI tests as the starting point, explores valid variations of their interactions, and generalizes the resulting task pattern into a reusable VA capability.

In this paper, we propose \toolname{}\footnote{\url{https://anonymous.4open.science/r/test2va-starter-kit-3261/README.md}}, an approach that reuses existing GUI tests to automatically construct app-specific VA capabilities.
A GUI test already provides grounded execution logic for a concrete user-facing task, but directly reusing the test limits the resulting capability to the specific task instance encoded in that test.
\toolname{} therefore aims to generalize this test-derived capability so that it can support broader task variations.

To discover such variations, \toolname{} adopts mutation as an exploration mechanism.
Instead of modifying application code or GUI structure, \toolname{} replaces individual test interactions with plausible alternatives, such as selecting a different GUI element or using a different input value.
These mutated interactions are treated as candidate task variations and are dynamically validated on the running application to determine whether they still preserve valid task behavior.
Given an application and one of its GUI tests, \toolname{} operates in three stages.
First, \textit{Test Interpretation and Context Collection} parses the original test into an executable interaction sequence, replays it on the target application, and collects the runtime GUI context associated with each interaction.
Second, \textit{Behavioral Variation Discovery and Validation} uses the original interactions and runtime context to generate mutation candidates, adapts affected assertions, and dynamically validates the resulting interaction sequences.
Finally, \textit{Capability Generalization} compares the original test with the validated variations to identify the common task pattern and construct a generalized VA capability.
Through this process, \toolname{} preserves the grounded task logic encoded in existing GUI tests while extending a concrete test instance to support broader user requests.

We evaluated \toolname{} on 100 Espresso GUI tests collected from ten open-source Android applications across different categories, covering 407 GUI events and 115 assertions.
For behavioral variation discovery, 396 events were successfully reproduced and used for mutation analysis.
Across these events, \toolname{} reduced 18,999 available mutation candidates to 1,272 predicted candidates, eliminating 93.31\% of the candidate space before dynamic validation.
Its assertion adaptation correctly identified 101 of 115 required event--assertion pairs, achieving 87.83\% accuracy.
After dynamic validation, \toolname{} identified mutable events with 83.33\% precision, 87.79\% recall, and an 85.50\% F1 score.
For capability generalization, \toolname{} generated 65 VA capabilities covering zero-, single-, and multi-mutable tasks.
Among them, 64 were successfully executed, yielding a 98.46\% execution success rate, and 56 satisfied our capability quality requirements.
We further analyzed the unsuccessful or low-quality cases to identify the main causes of failure across behavioral variation discovery, validation, and capability generation.
Dynamic validation required 127.71 seconds per event on average.

We further conducted a user study with 18 participants to evaluate whether the generated capabilities can support real app tasks through natural-language requests.
Across 150 task observations, the VA successfully completed 123 tasks, achieving an overall success rate of 82.0\%.
The success rates were 97.2\%, 81.2\%, and 55.6\% for zero-, single-, and multi-parameter capabilities, respectively.
In the post-study survey, 16 of 18 participants rated the VA as useful or very useful, providing additional evidence that test-derived app-specific capabilities can support practical interactions.

Overall, this paper makes the following contributions:
\begin{itemize}
    \item We propose \toolname{}, an approach that generalizes existing GUI tests into reusable app-specific VA capabilities by discovering and validating behavioral variations of the tested task.

    \item We evaluate \toolname{} on 100 GUI tests from ten real-world Android applications. Among 65 generated capabilities, 64 execute successfully and 56 satisfy our capability quality requirements.

    \item We analyze the unsuccessful and low-quality cases to identify common causes of errors in behavioral variation discovery, validation, and capability generation.

    \item We conduct a user study with 18 participants and 150 task observations to evaluate the generated capabilities in real app tasks. The VA successfully completes 82.0\% of the task observations, and 16 of 18 participants rate the VA as useful or very useful.
\end{itemize}

\section{Background} \label{sec:background}

The first three subsections present the background knowledge used to design and implement \toolname{}.
The last subsection explains the generated VA capability and how it can support VA development.

\subsection{Automated GUI Testing} \label{subsec:b_auto_gui_testing}

Most software, including mobile applications, is designed with graphical user interfaces (GUIs) containing visible components such as buttons, menus, and input boxes.
App testers write GUI test scripts that automatically execute testing tasks on these interfaces and verify whether application functions behave as expected.
A GUI test typically consists of a sequence of interactions followed by one or more assertions.
These scripts can be written manually or generated using recording services provided by testing frameworks.
For Android, Espresso~\cite{android_espresso} supports recording interactions and inserting assertions, while the platform-independent Appium~\cite{appium_tool} also supports interaction recording and automated execution.
GUI testing can also be automated using GUI agents.
Early agents relied on random~\cite{monkey_gui_testing} or rule-based~\cite{hellmann2011rule} strategies.
Later approaches incorporated Machine Learning, Natural Language Processing, and Computer Vision, while recent LLM-based agents further improve application-context interpretation and complex interaction execution.

Regardless of how a GUI test is produced, its interaction sequence usually represents one concrete execution of an application task.
For example, a test may configure a particular application setting, create a specific reminder, or select one particular item from a list.
Such a test therefore captures one concrete task instance that can later be analyzed to determine whether some of its interactions can support broader task variations.

\subsection{GUI Event and GUI Hierarchy} \label{subsec:b_gui}

Regardless of the automated GUI testing approach, a test execution can be represented as a sequence of \textit{GUI events}.
Each event consists of a \textit{GUI element} and a \textit{GUI control}; for example, in ``clicking a button,'' the button is the element and clicking is the control.
Some controls may also include an argument or value, such as text entered into an input field.
GUI elements contain attributes such as text, content description, bounds, and display status.
Testing frameworks use these attributes to locate target elements, while assertions examine element properties or application states to determine whether the expected behavior is observed.
Depending on the complexity of an application, a single screen may contain hundreds of GUI elements.
To organize these elements, mobile operating systems represent the current interface using a hierarchical structure, commonly referred to as a \textit{GUI hierarchy}. This hierarchy represents structural relationships among elements, such as parent, child, and sibling relationships, and may include both visible components and invisible layout containers. Because the hierarchy can be large and structurally complex, identifying plausible alternative elements for a GUI event is non-trivial. In \toolname{}, this structure provides the runtime context used to narrow the search for potential behavioral variations.

\subsection{GUI Event Mutation and Behavioral Variation} \label{subsec:b_mutant_gui}

Mutation testing~\cite{mutation} evaluates test effectiveness by introducing small changes, or \textit{mutations}, into the program. The resulting \textit{mutants} are then tested; those that still pass are referred to as \textit{surviving mutants}.
In this paper, we adopt mutation as an exploration mechanism for GUI interactions. Instead of modifying application code or GUI structures, \toolname{} mutates individual GUI events by replacing the target GUI element or GUI control with a plausible alternative.
This process is referred to as \textit{mutation generation}, where mutation operators define how an original event can be modified. Existing GUI mutation operators~\cite{GUIMutate} often change GUI properties or structures, such as shifting, modifying, or removing elements. In contrast, \toolname{} keeps the GUI structure unchanged and explores alternative interactions with the existing interface.
Each generated alternative is treated as a \textit{mutation candidate}. \toolname{} dynamically re-executes the modified event sequence and adapts the assertion when it depends on the mutated event. Candidates that execute successfully and satisfy the corresponding assertions are retained as \textit{surviving mutations}, representing validated behavioral variations of the original task.

\subsection{Voice Assistant Development} \label{subsec:b_av_developement}

A VA typically includes speech recognition, natural language understanding, text-to-speech, and task execution components.
In this paper, we focus on the \textit{task execution} component, which determines how a user-requested task is performed on the target application.
Task execution can be supported in different ways.
For example, agentic assistants may interpret the current application screen and dynamically determine the interactions needed to complete a request.
Alternatively, existing application knowledge can provide grounded execution logic.
\toolname{} follows the latter direction by reusing GUI tests, which already encode developer-defined, user-facing interaction sequences for concrete task instances.
A VA capability captures the execution logic of a task while exposing user-specific choices as parameters.
For example, a timer capability may use \texttt{setTimer(time\_length)}, where the method body implements the timer-setting interactions and \texttt{time\_length} represents the requested duration.
Similarly, \toolname{} generalizes the task pattern encoded by a GUI test and its validated behavioral variations into a reusable VA capability, preserving invariant interactions as execution logic while exposing validated variations as runtime parameters.

\section{Motivation Example} \label{sec:example}

\begin{figure*}[!htbp]
    \centering
     \includegraphics[width=0.98\linewidth]{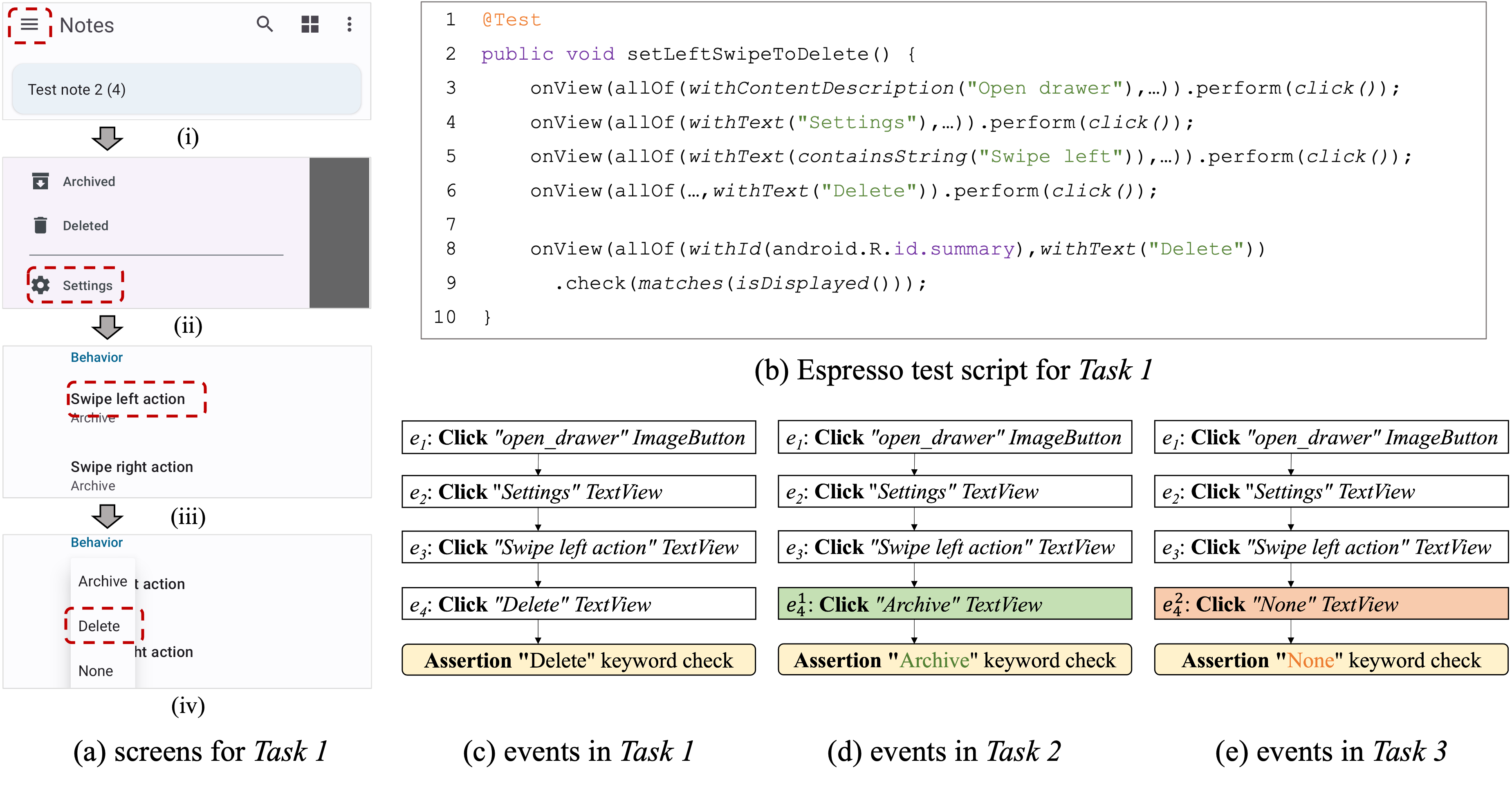}      
      \vspace{-0.3cm}
    \caption{Motivation Example of Android Application \textit{Another notes app}}   
    \label{fig:example}
    \vspace{-0.3cm}
\end{figure*}

Mobile applications provide many user-facing functions, such as setting an alarm in a clock application, playing a song in a music application, or configuring a gesture behavior in a note-taking application.
In this paper, we refer to such user-oriented application functionalities as \textit{tasks}.
A task is realized through a sequence of GUI interactions between the user and the application.
Developers and testers often encode such interaction sequences in GUI test scripts to verify that a specific application behavior is implemented correctly.
However, a GUI test typically captures only one concrete realization of a task.
In many cases, some interactions in that realization are required by the task structure, while others represent instance-specific choices that may vary across different usage scenarios.
This observation motivates us to distinguish between a concrete task instance encoded in a GUI test and the more general task pattern that the test may represent.

\textbf{\textit{Task Execution}.}
Figure~\ref{fig:example}(a) shows a sequence of four GUI interactions from the Android application \textit{Another Notes App}.
In screenshot (i), the user clicks the navigation drawer button highlighted by the red dashed box.
This opens the navigation menu shown in screenshot (ii), where the user selects ``Settings.''
The application then displays the settings page, as shown in screenshot (iii), where the user selects the ``Swipe left action'' option.
Finally, screenshot (iv) shows three available actions: ``Archive,'' ``Delete,'' and ``None.''
The user selects ``Delete.''
Together, these interactions perform a task that configures the application's swipe-left gesture to delete a note.

\textbf{\textit{Concrete Task Instance in a GUI Test}.}
Figure~\ref{fig:example}(b) shows how the same interaction sequence is encoded in a GUI test script.
The test first locates the navigation drawer button using its content description and invokes \texttt{click()} through \texttt{perform()}.
It then selects the ``Settings'' element, opens the ``Swipe left action'' setting, and clicks ``Delete.''
The test concludes with an assertion that checks whether the expected ``Delete'' value is displayed on the screen.
If the expected value is found, the test passes.
This test therefore represents one \textit{concrete task instance}: configuring the swipe-left action specifically to ``Delete.''
More generally, GUI tests frequently encode one particular choice within a broader application task.
For example, a test for calling a contact may invoke a call to ``Bob,'' even though the underlying task supported by the application is more generally to call an arbitrary contact from the contact list.
In such a case, ``Bob'' is specific to the concrete test instance rather than intrinsic to the general task itself.

\textbf{\textit{Generalized Task Pattern}.}
We next consider whether the concrete task instance can be generalized by modifying some of its interactions while preserving the underlying task.
Figure~\ref{fig:example}(c)--(e) shows the original task instance and two valid variations.
All three executions contain four interaction events followed by an assertion.
Events $e_1$--$e_3$ remain unchanged across the three executions:
the user opens the navigation drawer, selects ``Settings,'' and opens the ``Swipe left action'' setting.
In contrast, event $e_4$ differs across the three task instances.
The original task selects ``Delete,'' while the two variations select ``Archive'' and ``None,'' respectively.
The corresponding assertions are also updated to reflect the selected action.

These variations reveal a common task pattern:
\textit{configure the swipe-left gesture to a selected action}.
The first three interactions define the common structure required to reach the target setting, while the final interaction specifies which action is selected.
We therefore distinguish between two types of events in a GUI test.
A \textit{non-mutable event} represents an interaction that remains fixed across valid instances of the same task.
A \textit{mutable event} represents an instance-specific interaction that may be replaced by another valid alternative while preserving the general task semantics.

In the example shown in Figure~\ref{fig:example}, events $e_1$--$e_3$ are non-mutable, while $e_4$ is mutable.
The three concrete instances can therefore be viewed as realizations of the same generalized task pattern, where the selected swipe action is treated as a parameter.
To support this task through a VA method, the non-mutable interactions can be preserved while the mutable interaction is parameterized.
Importantly, not every GUI test necessarily contains mutable events.
Some tests may already represent tasks with no generalizable interaction choices, while others may contain one or multiple mutable events.
Therefore, the goal of \toolname{} is to determine whether valid task variations exist and, when they do, identify the interactions that can be parameterized.
This example motivates three key challenges in converting a concrete GUI test into a reusable VA capability:
(1) executing the original test and collecting the runtime context of the Task Instance to understand its variations,
(2) identifying and dynamically validating possible task variations, and
(3) extracting the common task pattern from the validated task instances and representing it as a generalized VA capability.

\section{Approach} \label{sec:app}

\subsection{Overview} \label{subsec:overview}

\begin{figure*}[!t]
    \centering
    \includegraphics[width=\textwidth]{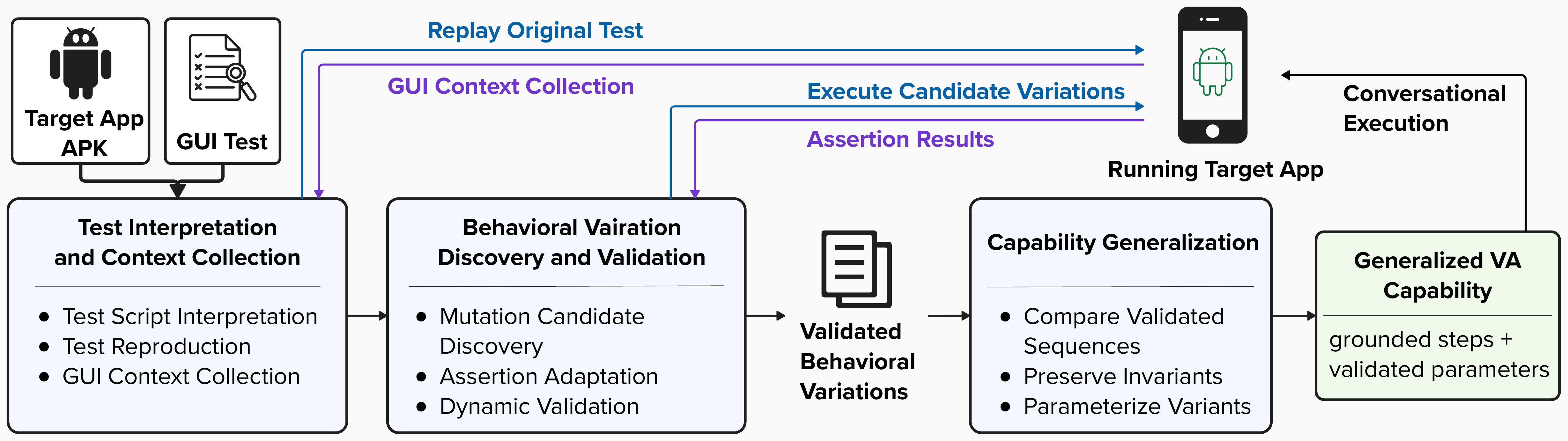}
    \caption{\toolname{} overview. Given a target app APK and a GUI test, \toolname{} first interprets the test and collects GUI context by replaying the original test on the running app. It then discovers mutation candidates, adapts assertions, and dynamically validates candidate variations. Finally, \toolname{} generalizes the validated behavioral variations into a VA capability with grounded execution steps and validated parameters.}
    \label{fig:overview}
    \vspace{-0.5cm}
\end{figure*}

\toolname{} reuses existing GUI tests to generate app-specific VA capabilities that preserve the original test logic while exposing validated behavioral variations as runtime parameters. As shown in Figure~\ref{fig:overview}, \toolname{} takes two inputs: the target app's APK file and its original GUI test script that performs a testing task. The output is a generalized VA capability that contains grounded execution steps and validated parameters, which can later be invoked through conversational execution in VA to support certain tasks.

The workflow has three conceptual stages. First, \toolname{} performs test interpretation and context collection. It interprets the test script, reproduces the original test on the running target app, and collects GUI context observed during execution. This stage grounds later reasoning in the app's actual runtime state rather than relying only on static test code.
Second, \toolname{} performs behavioral variation discovery and validation. Using the original test and collected GUI context, \toolname{} identifies mutation candidates for test events and adapts assertions when a candidate changes values checked by the original oracle. It then executes candidate variations on the running target app and collects assertion results. Candidates that execute successfully and pass the corresponding assertions are retained as validated behavioral variations.
Third, \toolname{} performs capability generalization. It compares the original test sequence with the validated variation sequences to determine which interactions remain invariant and which interactions can vary. The invariant interactions become the grounded execution logic of the generated capability, while validated variants are parameterized so that the capability can support related task requests at runtime.

Through this workflow, \toolname{} generalizes a concrete GUI test into a capability that supports a broader set of related VA requests. The original test provides grounded execution logic, while dynamic validation determines which behavioral variations can be safely exposed as parameters. The generated capability therefore preserves test-derived execution steps while allowing users to instantiate broader task variants through conversational requests.

\subsection{Test Interpretation and Context Collection}
\label{subsec:test_interpretation}

Before generalizing a test-derived capability, \toolname{} first interprets the original GUI test and collects the application context associated with its execution.
These data will be used to identify and validate potential behavioral variations of the original test interactions
in Section~\ref{subsec:capability_generalization}.

\subsubsection{Test Script Interpretation}

The Test Script Parser interprets the original test code (e.g., an Espresso test) into an execution notation that can be consumed by the subsequent variation discovery and validation processes.
This notation enables \toolname{} to follow a framework-independent design, allowing the same test-reuse approach to be extended to different platforms and testing frameworks.
To construct the notation, the Parser adopts srcML~\cite{srcml}, a multi-language parsing tool that converts source code into XML for structural analysis, and extracts the behavioral information needed to reproduce and analyze the tested interactions.
This information includes the test setup and cleanup procedures, GUI interaction sequence, target GUI elements and their matchers, actions and corresponding arguments or values, and assertions.
In particular, the extracted GUI interactions form an event sequence $E=\langle e_1,e_2,\ldots,e_n\rangle$, where each event $e_i$ records its target GUI element, action, and corresponding arguments, if any.
Together with the setup, cleanup, and assertion information, this structured notation preserves the original test logic and provides the input for test reproduction.

\subsubsection{Test Reproduction and Context Collection}

The Base Test Reproducer re-executes the parsed interaction sequence to collect the runtime application context needed for behavioral variation discovery.
While the test code specifies the GUI elements and actions exercised by the developer, identifying potential variations also requires understanding what alternative interactions are available at each point during execution.
To reproduce the test, \toolname{} uses the Appium testing framework to map common GUI actions in the execution notation, such as clicking, text input, and swiping, to corresponding executable operations on the target application.

During reproduction, \toolname{} uses Android Debug Bridge (adb)~\cite{android_adb} and UIAutomator~\cite{android_ui_automator} to collect the GUI hierarchy before and after each event.
For an interaction sequence $E=\langle e_1,e_2,\ldots,e_n\rangle$, the collected runtime contexts are represented as $C=\langle c_0,c_1,\ldots,c_n\rangle$, where $c_0$ represents the GUI context before $e_1$, and each $c_i$ represents the context after executing $e_i$.
Together, each interaction $e_i$ and its preceding context $c_{i-1}$ provide the behavioral and GUI information needed to identify plausible variations of $e_i$.
Building on the extracted interactions and runtime contexts, \toolname{} next explores potential behavioral variations for each event and dynamically validates whether these variations preserve the intended task behavior.

\subsection{Behavioral Variation Discovery and Validation}
\label{subsec:behavioral_variation}

To systematically explore behavioral variations, \toolname{} adopts mutation as an exploration mechanism by replacing GUI events with plausible alternatives. Unlike traditional mutation testing, our goal is to identify \textit{surviving mutations} that can be successfully executed and satisfy the corresponding assertions; these survivors represent valid behavioral variations of the original task. Rather than exhaustively mutating each event with all possible interactions, which would create a large and mostly irrelevant search space, \toolname{} selects a limited set of \textit{mutation candidates} that are more likely to preserve the underlying task behavior.

To identify these candidates, \toolname{} considers the two main parts of a GUI event: the target GUI element and the GUI control applied to it. Accordingly, a mutation may replace the target element, modify the control or its value, or combine both. For each event $e_i$, \toolname{} uses the event together with its preceding runtime context $c_{i-1}$ to identify plausible candidates. These candidates are then dynamically validated to determine which mutations survive and can be treated as behavioral variations.

\subsubsection{Mutation Candidate Discovery}
\label{subsubsec:mutation_candidate}

\toolname{} identifies mutation candidates separately for GUI controls and GUI elements, as each requires different strategies for generating plausible alternatives.

\paragraph{GUI Control Mutation Candidates.}
Mutation rules for GUI controls depend on the type of the original action.
Since our objective is to discover surviving mutations, a mutated control 
should generally preserve the interaction semantics of its original event.
For instance, the mutation of clicking a button is more likely to involve 
clicking an alternative element than replacing the click with a text-editing 
operation.
Based on this observation, we constrain control mutations according to their 
interaction types, as summarized in Table~\ref{tab:mutant_operator_control}.

\begin{table}[]
\caption{\toolname{} GUI Control Mutation Candidate Rules}
\label{tab:mutant_operator_control}
\begin{tabular}{|c|c|l|l|l|}
\hline
\# & \multicolumn{1}{l|}{\textbf{Control Type}} 
   & \textbf{Original Control} 
   & \textbf{Mutated Control} 
   & \textbf{Applicable GUI Components} \\ \hline

1 & \multirow{3}{*}{Click}
  & click() 
  & click() 
  & \multirow{3}{*}{\begin{tabular}[c]{@{}l@{}}
      Button, TextView, image,\\
      switch, checkbox
    \end{tabular}} \\ \cline{1-1} \cline{3-4}

2 & 
  & longClick() 
  & longClick() 
  & \\ \cline{1-1} \cline{3-4}

3 & 
  & doubleClick() 
  & doubleClick() 
  & \\ \hline

4 & \multirow{3}{*}{Type In}
  & replaceText($c_1$) 
  & replaceText($c_2$) 
  & \multirow{3}{*}{\begin{tabular}[c]{@{}l@{}}
      Input fields such as\\
      EditText
    \end{tabular}} \\ \cline{1-1} \cline{3-4}

5 &
  & typeText($c_1$)
  & typeText($c_2$)
  & \\ \cline{1-1} \cline{3-4}

6 &
  & clearText()
  & clearText()
  & \\ \hline

7 & Scroll
  & scrollTo()
  & scrollTo()
  & \begin{tabular}[c]{@{}l@{}}
      RecyclerView, Spinner,\\
      ScrollView
    \end{tabular} \\ \hline

8 & \multirow{2}{*}{Swipe}
  & swipeLeft()
  & swipeRight()
  & \multirow{2}{*}{\begin{tabular}[c]{@{}l@{}}
      ProgressBar and child\\
      views in a list
    \end{tabular}} \\ \cline{1-1} \cline{3-4}

9 &
  & swipeRight()
  & swipeLeft()
  & \\ \hline

10 & System
   & pressBack()
   & pressBack()
   & System navigation back \\ \hline

\end{tabular}
\end{table}

For parameterized controls, the mutation may modify the control value while 
preserving its semantic type.
For example, given an input action with the original value ``mushroom,'' 
\toolname{} uses the original value together with the application context to 
identify its semantic type (e.g., a food name) and generate alternative values 
such as ``apple,'' ``broccoli,'' or ``pasta.''
Similarly, directional actions such as swiping may be mutated to an alternative 
direction when supported by the control mutation rules.
In this way, the rules constrain how an event can be mutated, while 
the application context helps determine meaningful candidate values.

\paragraph{GUI Element Mutation Candidates.}
For GUI elements, \toolname{} aims to identify alternative elements that may support the same interaction while representing different user choices. Instead of considering all elements on the screen, \toolname{} first uses structural similarity in the GUI hierarchy to narrow the candidate space. Elements providing alternative choices are often grouped within the same layout and therefore appear as sibling nodes in the hierarchy.

\begin{figure*}[t]
    \centering
    \includegraphics[width=\textwidth]{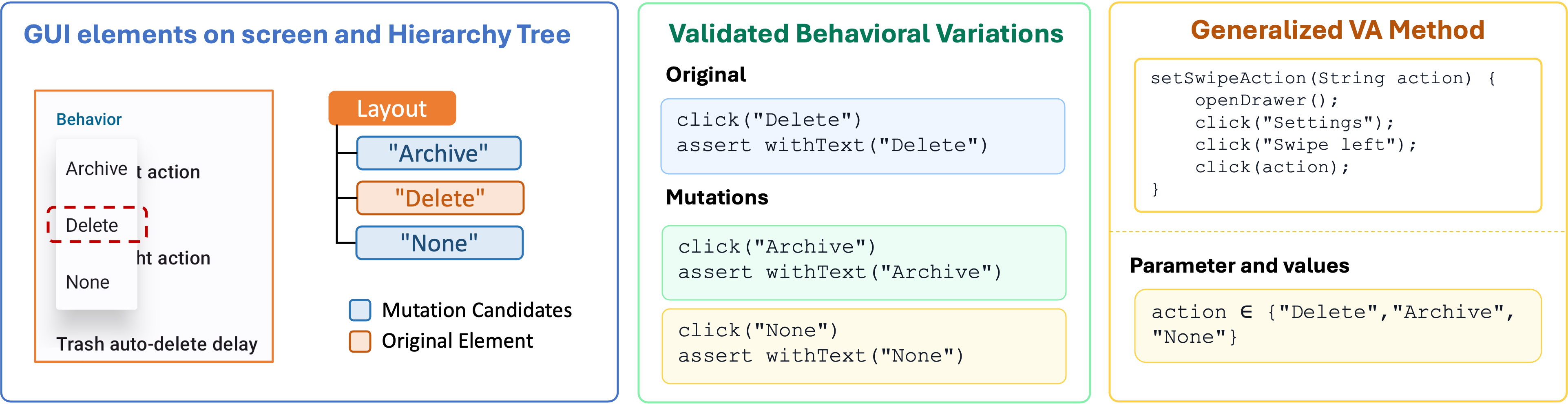}
    \caption{Running example of behavioral variation discovery and capability generalization. 
    \toolname{} identifies ``Archive'' and ``None'' as mutation candidates for the original ``Delete'' element based on the GUI hierarchy (left), dynamically validates them as behavioral variations with adapted assertions (middle), and generalizes the validated variations into a parameterized VA method (right).}
    \label{fig:approach_example}
    \vspace{-0.5cm}
\end{figure*}

Figure~\ref{fig:approach_example} illustrates this process using the ``Delete'' element from the original interaction. As shown in the left panel, ``Archive,'' ``Delete,'' and ``None'' are sibling elements under the same layout. \toolname{} therefore identifies ``Archive'' and ``None'' as mutation candidates for the original ``Delete'' element. These candidates are subsequently executed and validated; as shown in the middle panel, both candidates survive and are identified as behavioral variations of the original interaction.

Structural similarity alone may miss plausible candidates because the GUI hierarchy captures structural relationships but not necessarily the semantic roles or functionality of the elements. To complement the structural rules, \toolname{} leverages an LLM to interpret both the test code and GUI context. For each event $e_i$, we provide its original Espresso statement and preceding GUI hierarchy $c_{i-1}$ to identify additional elements that may serve similar functionality. We restrict the candidates to the same GUI class as the original element to ensure that the original action remains applicable. The LLM-discovered candidates supplement those identified through structural similarity, providing broader coverage while maintaining a constrained mutation space.

\paragraph{Assertion Adaptation.}

Dynamic validation uses test assertions to determine whether a mutation preserves valid task behavior. However, assertions may depend on the concrete values of the original event, causing valid mutations to fail due to outdated assertions. To avoid such false rejections, \toolname{} uses an LLM to identify assertions affected by the mutated event and adapt their values accordingly, while leaving unrelated assertions unchanged. Figure~\ref{fig:approach_example} illustrates this process, where the assertions are adapted along with the ``Archive'' and ``None'' mutations.

\subsubsection{Dynamic Validation}
\label{subsubsec:dynamic_validation}

The mutation candidates represent plausible alternatives to the original events, but they do not necessarily preserve valid task behavior. Therefore, \toolname{} dynamically validates each candidate on the target application to identify \textit{surviving mutations}. For each event $e_i$, \toolname{} evaluates its mutation candidates individually. Before evaluating a candidate, the application is reset to its initial state, and the original test setup is reproduced. \toolname{} then executes the original event sequence, replacing only $e_i$ with the candidate while keeping the remaining interactions unchanged.

A candidate is rejected if the resulting sequence encounters an execution error or fails its corresponding assertions. Otherwise, the candidate is recorded as a \textit{surviving mutation}. When an assertion depends on the mutated event, \toolname{} uses the adapted assertion described above during validation. In this way, each surviving mutation represents an executable alternative to the original interaction and is treated as a \textit{behavioral variation} of the test-derived capability.

Figure~\ref{fig:approach_example} illustrates this process. For the original ``Delete'' interaction, both ``Archive'' and ``None'' successfully execute with their adapted assertions and are therefore retained as behavioral variations. Together with the original interaction, these validated variations provide multiple task instances from which \toolname{} generalizes the VA capability in the next stage.

\subsection{Capability Generalization}
\label{subsec:capability_generalization}

After identifying the validated behavioral variations, \toolname{} generalizes them into a reusable VA capability. Specifically, \toolname{} compares the original event sequence with its surviving mutation sequences to identify the interactions that remain invariant and those that vary across the validated task instances. Interactions without surviving mutations are treated as invariant and preserved in the generated capability with their original GUI actions and element features. In contrast, interactions with surviving mutations are treated as variable, and their differences are extracted and parameterized. In this way, the generated capability preserves the common interaction pattern encoded by the original test while exposing validated variations as configurable parameters.

Figure~\ref{fig:approach_example} illustrates this generalization using the running example. The interactions for opening the drawer, navigating to ``Settings,'' and selecting ``Swipe left'' remain unchanged across the original and mutated sequences and are therefore preserved in the generated VA method. The subsequent interaction, however, has three validated instances: ``Delete'' from the original test and the surviving ``Archive'' and ``None'' mutations. Since these instances differ only in the target option, \toolname{} parameterizes this difference as \texttt{action}, producing a single method that supports all three validated values. Thus, rather than generating separate methods for each surviving sequence, \toolname{} captures their common task pattern in one generalized capability.

The parameterization strategy depends on the source of the varying value. Some GUI elements represent developer-defined choices, such as ``Delete,'' ``Archive,'' and ``None'' in Figure~\ref{fig:approach_example}. Because these values are predefined by the application, the generated parameter is constrained to the validated alternatives so that the corresponding GUI element can be correctly located at runtime. Other interactions contain values created by users, such as a contact name, reminder date, or to-do item. Such values are not limited to those observed during mutation and can instead be directly parameterized to accept runtime input. \toolname{} leverages the LLM's interpretation of the original and mutated statements to distinguish developer-defined alternatives from user-provided values and applies the corresponding parameterization strategy.

Through this process, \toolname{} generalizes a concrete GUI test into a capability that captures a broader family of behaviorally validated task instances. Invariant interactions retain the grounded execution logic provided by the original test, while variable interactions expose the flexibility discovered through mutation and dynamic validation. Consequently, instead of reusing one GUI test to support only the concrete behavior encoded in that test, \toolname{} enables the same test logic to support multiple related VA requests through a single generalized capability.

\section{Evaluation}
\label{sec:evaluation}

We evaluate \toolname{} on GUI tests collected from open-source Android applications. The evaluation focuses on whether \toolname{} can discover valid behavioral variations from existing GUI tests, generalize the validated variations into reusable VA capabilities, and perform this process with practical offline cost. We answer the following research questions:

\begin{itemize}[leftmargin=*]
    \item \textbf{RQ1:} How effectively does \toolname{} generate mutation candidates and adapt test assertions?
    \item \textbf{RQ2:} How accurately and efficiently does \toolname{} validate behavioral variations at runtime?
    \item \textbf{RQ3:} How effectively does \toolname{} generate executable and reusable VA capabilities from correctly validated GUI tests?
    \item \textbf{RQ4:} How effectively can users use the generated VA capabilities to complete real-app tasks?
\end{itemize}

\subsection{Experiment Setup}
\label{subsec:eval_setup}

We evaluate \toolname{} on 100 Espresso GUI tests collected from ten open-source Android applications across different categories.
Our dataset extends the evaluation dataset used in AppVA~\cite{qin2025apptova}.
We reuse its 20 GUI tests from five applications unchanged and add 80 new GUI tests from five additional applications.
The applications were selected from F-Droid~\cite{fdroid}; we excluded games, applications not written in Java or Kotlin, and applications that did not support Espresso test recording. For applications with existing UI tests, we reused their test scenarios and replaced mocking-dependent parts with recorded Espresso interactions. For applications without UI tests, we recorded tests for commonly used app functionalities. Each application contributes ten test cases, and each test starts from the application's main activity so that the resulting capability matches the normal starting point of VA task execution. In total, the dataset contains 100 GUI test methods, including 407 GUI events and 115 assertions, with event-sequence lengths ranging from 1 to 11 and assertion counts ranging from 1 to 6.
We run test reproduction and mutation validation on an Android emulator using Android 13.0, and execute generated VA capabilities on a Pixel 5 device running Android 11. For LLM-based steps, including mutation candidate generation, assertion adaptation, and capability generation, we use GPT-4o with default model parameters. 

\subsection{RQ1: Mutation Candidate Generation and Assertion Adaptation}
\label{subsec:rq1}

RQ1 evaluates how effectively \toolname{} prepares behavioral variations for runtime validation. 
This stage consists of two steps: generating a constrained set of mutation candidates for each GUI event and adapting test assertions when their expected values depend on the mutated event.
We evaluate candidate generation by measuring how much \toolname{} reduces the available GUI candidate space, and evaluate assertion adaptation by comparing the predicted event--assertion relationships against the ground truth.

\paragraph{Mutation candidate generation.}
A brute-force mutation strategy could consider all available GUI elements on each screen as possible alternatives to the original event.
However, many of these candidates are unrelated to the intended task and would unnecessarily increase the cost of subsequent runtime validation.
\toolname{} therefore applies GUI-structure rules and LLM-based semantic filtering to reduce the candidate space.

Table~\ref{tab:rq1_candidate_generation} reports the number of available mutation candidates before and after filtering.
Although the dataset contains 407 GUI events in total, 11 events were excluded from runtime validation because the original tests could not be reliably reproduced:
eight events from Plees Tracker (v24.8.1) and three events from PocketPlan (v1.4.2). This leaves 396 events for mutation analysis.
Across these 396 reproduced events, \toolname{} reduces the candidate space from 18,999 available candidates to 1,272 predicted candidates, eliminating 93.31\% of candidates before runtime validation.
This reduction substantially limits the number of candidate executions required in the next stage while retaining plausible behavioral variations for validation.

\begin{table}[t]
\centering
\caption{Mutation candidate generation results.}
\label{tab:rq1_candidate_generation}
\begin{tabular}{lrrr}
\toprule
\textbf{App} & \textbf{Events} & \textbf{Before Filtering} & \textbf{After Filtering} \\
\midrule
Another Notes App & 32 & 1210 & 107 \\
Broccoli & 43 & 1864 & 121 \\
Diaguard & 52 & 2986 & 196 \\
MedTimer & 46 & 2125 & 121 \\
MyExpenses & 36 & 1867 & 122 \\
Plees Tracker & 31 & 1451 & 177 \\
PocketPlan & 34 & 1911 & 87 \\
Calendar & 33 & 1384 & 73 \\
A Flash Deck & 45 & 1419 & 114 \\
OpenWorkout & 44 & 2782 & 154 \\
\midrule
\textbf{Total} & \textbf{396} & \textbf{18,999} & \textbf{1,272} \\
\bottomrule
\end{tabular}
\end{table}

\paragraph{Assertion adaptation.}
Runtime validation relies on the assertions from the original GUI test to determine whether a mutation preserves the intended task outcome.
When a mutated event changes a value referenced by an assertion, the corresponding assertion must also be adapted; otherwise, a valid behavioral variation may be incorrectly rejected.
We evaluate assertion adaptation as a pairwise prediction problem.
Each prediction identifies whether an assertion should be updated for a target event, including cases where no assertion modification is required.
A prediction is considered correct only when the predicted event--assertion relationship matches the ground truth.
As shown in Table~\ref{tab:assertion_adaptation}, \toolname{} correctly identifies 101 of 115 required event--assertion relationships, achieving an accuracy of 87.83\%.
Together, the candidate filtering and assertion adaptation results show that \toolname{} substantially reduces the behavioral variation search space while preparing most candidate mutations with the appropriate validation assertions for subsequent runtime validation.

\begin{table}[t]
\centering
\caption{Assertion adaptation accuracy.}
\label{tab:assertion_adaptation}
\begin{tabular}{lrr}
\toprule
\textbf{App} & \textbf{Required pairs} & \textbf{Correct pairs} \\
\midrule
Another Notes App & 11 & 11 \\
Broccoli & 12 & 12 \\
Diaguard & 12 & 12 \\
MedTimer & 13 & 9 \\
MyExpenses & 11 & 11 \\
Plees Tracker & 14 & 8 \\
PocketPlan & 10 & 9 \\
Calendar & 10 & 8 \\
A Flash Deck & 12 & 12 \\
OpenWorkout & 10 & 9 \\
\midrule
\textbf{Total} & \textbf{115} & \textbf{101} \\
\bottomrule
\end{tabular}
\end{table}

\subsection{RQ2: Runtime Validation of Behavioral Variations}
\label{subsec:rq2}

RQ2 evaluates how accurately and efficiently \toolname{} validates the behavioral variations prepared in RQ1.
For each event, \toolname{} executes the predicted mutation candidates together with the adapted assertions and determines whether the event is mutable.
We evaluate this stage from two perspectives: validation accuracy and runtime cost.

\paragraph{Validation accuracy.}
After dynamic validation, each event is classified as mutable or non-mutable.
We evaluate this classification against the ground truth at the event level.
A true positive (TP) is a mutable event correctly identified as mutable, while a true negative (TN) is a non-mutable event correctly identified as non-mutable.
A false positive (FP) is a non-mutable event incorrectly identified as mutable, and a false negative (FN) is a mutable event incorrectly identified as non-mutable.
As shown in Table~\ref{tab:rq2_event_identification}, \toolname{} correctly identifies 115 mutable events and 242 non-mutable events.
Based on these results, \toolname{} achieves 83.33\% precision, 87.79\% recall, and an F1 score of 85.50\% for mutable-event identification.
The errors include both false positives and false negatives.
False positives may expose unsupported behavioral variations to later capability generation, while false negatives may cause valid reuse opportunities to be missed.
Therefore, for the capability-generation evaluation, we retain only test methods whose mutable and non-mutable events are all correctly identified according to the ground truth.
This allows us to evaluate capability generation independently from errors introduced during runtime validation.

\begin{table}[t]
\centering
\caption{Mutable and non-mutable event identification results.}
\label{tab:rq2_event_identification}
\begin{tabular}{lrrrr}
\toprule
\textbf{App} & \textbf{TN} & \textbf{TP} & \textbf{FN} & \textbf{FP} \\
\midrule
Another Notes App & 21 & 9  & 0 & 2 \\
Broccoli          & 20 & 21 & 0 & 2 \\
Diaguard          & 35 & 13 & 0 & 4 \\
MedTimer          & 25 & 15 & 4 & 2 \\
MyExpenses        & 25 & 9  & 0 & 2 \\
Plees Tracker     & 22 & 3  & 6 & 0 \\
PocketPlan        & 22 & 8  & 0 & 4 \\
Calendar          & 16 & 11 & 3 & 3 \\
A Flash Deck      & 26 & 14 & 3 & 2 \\
OpenWorkout       & 30 & 12 & 0 & 2 \\
\midrule
\textbf{Total} & \textbf{242} & \textbf{115} & \textbf{16} & \textbf{23} \\
\bottomrule
\end{tabular}
\end{table}

\paragraph{Runtime validation cost.}
Dynamic validation requires executing the predicted candidates on the target application and is therefore the main runtime cost of the offline generation process.
Table~\ref{tab:rq2_validation_cost} reports the validation time for each application.
Across the 396 reproduced events, dynamic validation takes an average of 127.71 seconds per event.
The cost varies across applications because each event may have a different number of predicted candidates, assertions, and execution paths.
Because this process is performed offline, the validation overhead does not affect runtime VA interaction.
Overall, the runtime validator correctly classifies most mutable and non-mutable events while keeping validation practical for offline capability generation.
Although validation accuracy is measured at the event level, capability generation operates at the test-method level.
We therefore consider a test method correctly validated only when all of its events are correctly classified as mutable or non-mutable.
Based on this criterion, 65 of 100 test methods are correctly validated and used as input to RQ3.

\begin{table}[t]
\centering
\caption{Dynamic validation cost by application.}
\label{tab:rq2_validation_cost}
\begin{tabular}{lrrr}
\toprule
\textbf{App} & \textbf{Events} & \textbf{Validation time (s)} & \textbf{Time/event (s)} \\
\midrule
Another Notes App & 32 & 3558.10 & 111.19 \\
Broccoli & 43 & 5005.00 & 116.40 \\
Diaguard & 52 & 5974.10 & 114.89 \\
MedTimer & 46 & 6094.20 & 132.48 \\
MyExpenses & 36 & 5303.90 & 147.33 \\
Plees Tracker & 31 & 7444.00 & 240.13 \\
PocketPlan & 34 & 3094.20 & 91.01 \\
Calendar & 33 & 2872.30 & 87.04 \\
A Flash Deck & 45 & 8653.00 & 192.29 \\
OpenWorkout & 44 & 2576.00 & 58.55 \\
\midrule
\textbf{Total / Average} & \textbf{396} & \textbf{50,574.80} & \textbf{127.71} \\
\bottomrule
\end{tabular}
\end{table}

\subsection{RQ3: VA Capability Generation}
\label{subsec:rq3}

RQ3 evaluates how effectively \toolname{} generates executable and reusable VA capabilities from correctly validated GUI tests.
As described in RQ2, runtime validation is measured at the event level, whereas capability generation operates at the test-method level.
We therefore evaluate the generated capabilities from the 65 test methods whose events are all correctly classified as mutable or non-mutable.
Not every correctly validated test should be generalized in the same way.
Tests with no mutable events should preserve the original behavior without introducing parameters, while tests with one or more mutable events should expose those variations as reusable parameters.
We therefore evaluate the generated capabilities according to their ground-truth generalizability.

\begin{table}[t]
\centering
\caption{Capability execution and quality by generalizability.}
\label{tab:rq3_capability_quality}
\begin{threeparttable}
\begin{tabular}{lrrrrrr}
\toprule
\textbf{Type} & \textbf{Total} & \textbf{Exec.} & \textbf{Usable} & \textbf{Execution err.} & \textbf{Code-quality err.} & \textbf{Generality err.} \\
\midrule
Zero   & 17 & 16 & 16 & 1 & 0 & 0 \\
Single & 34 & 34 & 31 & 0 & 1 & 2 \\
Multi  & 14 & 14 & 9  & 0 & 1 & 5 \\
\midrule
\textbf{Total} & \textbf{65} & \textbf{64} & \textbf{56} & \textbf{1} & \textbf{2} & \textbf{7} \\
\bottomrule
\end{tabular}
\end{threeparttable}
\end{table}

We manually classify the 65 generated capabilities into three categories based on the number of mutable events identified in the corresponding test method.
A zero-mutable capability contains no validated behavioral variation and should preserve the original concrete behavior.
A single-mutable capability contains one mutable event and should expose the corresponding variation as a parameter.
A multi-mutable capability contains multiple mutable events and should preserve the invariant interaction sequence while parameterizing multiple variable interactions.
As shown in Table~\ref{tab:rq3_capability_quality}, \toolname{} successfully executes 64 of the 65 generated capabilities, yielding a 98.46\% execution success rate.
This result shows that the generated capabilities generally preserve the executable behavior of the original GUI tests.
However, successful execution alone does not indicate whether a generated capability is appropriately generalized.
We therefore manually inspect each capability and evaluate whether its parameters and generated code correctly represent the validated behavioral variations.
Overall, 56 of the 65 capabilities satisfy the usability and generalization requirements.

The observed errors fall into three categories:
1) Execution errors occur when a generated capability cannot successfully complete the intended task at runtime;
2) Code-quality errors occur when the generated capability uses poorly designed or difficult-to-interpret parameter representations, such as unreadable GUI identifiers or improperly formatted predefined alternatives.
3) Generality errors occur when a capability executes successfully but supports only a limited set of values even though the corresponding task should allow broader user-provided inputs.
The results also show that generalization becomes more challenging as the number of mutable events increases.
All 16 successfully executed zero-mutable capabilities satisfy the quality requirement, while 31 of 34 single-mutable capabilities and 9 of 14 multi-mutable capabilities are considered usable.
Most errors in the multi-mutable category are generality errors. In these cases, the generated capability works for the tested values but does not generalize correctly to other valid user inputs. For example, the generated method may hard-code a small set of observed alternatives instead of allowing a user-provided value.

Overall, \toolname{} can generate executable VA capabilities from correctly validated GUI tests with high execution success, while the remaining errors primarily arise from the quality of parameterization and generalization rather than from loss of the original executable behavior.

\subsection{RQ4: User Study on Capability Use}
\label{subsec:rq4}

RQ4 evaluates how effectively users can use the generated VA capabilities to complete real application tasks.
Unlike RQ3, which evaluates the executability and generalization quality of the generated capabilities themselves, RQ4 examines their use in an end-to-end VA setting where users express task requests in natural language and the VA invokes the corresponding capability.

\paragraph{Study design.}
We conduct a user study using the capabilities that satisfy the usability and generalization requirements in RQ3.
Among the 56 usable generated capabilities, we randomly select five capabilities from each of the ten applications, and get a total number of 50 capabilities.
Each selected capability defines one study task.
Each task contains two phases.
First, the participant completes the task manually by interacting with the application.
Second, the participant uses the VA to complete a related task supported by the selected capability.
The VA task may use different concrete values from the manual task, allowing us to examine whether parameterized capabilities can support runtime user inputs rather than only reproducing the original test values.

We initially recruited 15 participants, with each participant completing 10 tasks.
Each of the 50 tasks was independently performed by three participants, resulting in 150 task observations.
When an observation did not follow the study protocol and could not be reliably evaluated, we excluded that observation and recruited three additional participants to repeat the corresponding tasks.
The final analysis therefore contains 150 valid observations, with three observations for each of the 50 tasks.
For the user study, we classify capabilities by the actual number of parameters that users must provide when invoking the generated capability: 12 zero-parameter, 32 single-parameter, and 6 multi-parameter tasks. 
This classification follows the generalization structure in RQ3, where zero-mutable capabilities typically require no runtime input, while single- and multi-mutable capabilities expose one or more runtime parameters.

\begin{table}[t]
\centering
\caption{User-study outcomes by capability parameterization.}
\label{tab:rq4_outcomes}
\begin{tabular}{lrrrrr}
\toprule
\textbf{Parameterization} & \textbf{Obs.} & \textbf{VA success} & \textbf{C3} & \textbf{C4} & \textbf{Succ. rate} \\
\midrule
Zero-param   & 36 & 35 & 1  & 0 & 97.2\% \\
Single-param & 96 & 78 & 14 & 4 & 81.2\% \\
Multi-param  & 18 & 10 & 6  & 2 & 55.6\% \\
\midrule
\textbf{Total} & \textbf{150} & \textbf{123} & \textbf{21} & \textbf{6} & \textbf{82.0\%} \\
\bottomrule
\end{tabular}
\end{table}

\paragraph{Task outcomes.}
We classify each observation according to whether the participant and the VA successfully complete the corresponding task.
C1 indicates that the participant does not successfully complete the manual task, while the VA completes the VA task.
C2 indicates that both the participant and the VA successfully complete their tasks.
C3 indicates that the participant completes the manual task, but the VA fails or encounters an issue.
C4 captures other outcomes that do not fit the previous cases.
We count C1 and C2 as VA success because the VA successfully completes the requested task in both cases.
As shown in Table~\ref{tab:rq4_outcomes}, the VA successfully completes 123 of the 150 task observations, yielding an overall success rate of 82.0\%.
Zero-parameter capabilities achieve 35 successful executions out of 36 observations (97.2\%), while single-parameter capabilities complete 78 of 96 observations (81.2\%).
Multi-parameter capabilities complete 10 of 18 observations (55.6\%).

The results show that the generated capabilities can support users in completing real application tasks, including tasks that require runtime values.
The lower success rate for multi-parameter tasks may reflect the greater end-to-end interaction complexity of collecting multiple user inputs, binding them to the correct capability parameters, and executing the task in the expected app state.
Because all capabilities included in the user study already satisfy the execution and usability requirements in RQ3, these failures are not necessarily caused by the executability of the generated capability itself.

\paragraph{User feedback.}
We further use the post-study survey to contextualize the task outcomes.
Sixteen of the 18 participants rated the VA as useful or very useful, and 11 of the 18 participants reported that they would be likely or very likely to use similar VA support in other applications, while four were neutral.
These responses suggest that participants generally found the generated app-specific VA capabilities useful for completing representative application tasks.
These results also support the motivation for \toolname{} by showing that users can benefit from app-specific VA capabilities generated from existing GUI tests. By reusing existing test artifacts rather than implementing each capability manually, \toolname{} has the potential to reduce the engineering effort required to provide such support across applications.

\section{Discussion and Limitations} \label{sec:discussion}

\paragraph{Scope and external validity.}
\toolname{} focuses on reusing app-level GUI tests to generate VA capabilities for application-specific tasks. It does not aim to cover system-level actions, such as changing Wi-Fi settings or modifying notification permissions, which are typically handled by operating-system accessibility tools such as Voice Access~\cite{voiceAccess} and VoiceOver~\cite{VoiceOver}. This scope allows \toolname{} to focus on custom app behaviors that are less likely to be covered by general-purpose OS-level assistants.
Our evaluation uses open-source Android applications because test construction, instrumentation, and capability validation require controllable app builds. This selection may not represent all commercial Android applications, especially apps with custom UI frameworks, unstable layouts, or restricted automation support. The user-study tasks also cover a bounded set of app functions and parameterization patterns. Although the tasks include zero-, single-, and multi-parameter capabilities, they do not exhaust all forms of app interaction. Future evaluations should include broader app categories, larger task sets, and tasks with richer conversational context.

\paragraph{Mutation search scope.}
\toolname{} currently searches for mutation candidates on the same app screen as the original event and treats each mutation as a replacement for a single GUI event. This design keeps the search and dynamic validation tractable, but it may miss valid variations where a user reaches the same state through a different navigation path or where one conceptual action corresponds to multiple low-level GUI events. A possible extension is to represent task logic using macro events~\cite{macro_GUIevents}, where a high-level action such as selecting an item from a menu can be modeled as one conceptual step even if it requires multiple GUI operations. This would allow \toolname{} to explore broader behavioral variations, but would also require more complex assertion adaptation and validation.

\paragraph{Pipeline error propagation.}
RQ3 evaluates capability generation using test methods whose mutable and non-mutable events are correctly identified according to the ground truth. This design allows us to evaluate capability generation independently from errors introduced during runtime validation. In the complete \toolname{} pipeline, however, false-positive or false-negative validation results may propagate into the generated capability. The end-to-end quality of automatically generated capabilities therefore depends on both the accuracy of behavioral variation validation and the quality of capability generalization.

\paragraph{End-to-end execution boundary.}
RQ4 evaluates generated capabilities in an end-to-end VA setting, where the generated capability is only one part of the execution path. A user request must be interpreted, its arguments must be extracted and bound to capability parameters, and the app must be in the expected state when the capability runs. This boundary is particularly relevant for multi-parameter tasks, where failures may arise from request interpretation, argument binding, or app-state mismatches even when the generated capability itself is executable. RQ4 therefore evaluates whether generated capabilities can support user-facing VA tasks in realistic app settings, while RQ1--RQ3 evaluate the generation pipeline itself.

\paragraph{Dependence on AI models.}
\toolname{} uses GPT-4o for mutation candidate selection, assertion adaptation, and capability generation. Better AI models or task-specific fine-tuning may improve candidate selection, assertion-event dependency detection, and parameter classification. However, these improvements would not change the central idea of the paper: existing GUI tests can be reused to discover behavioral variations, validate those variations dynamically, and generalize the validated behavior into VA capabilities. The model-dependent components primarily affect the accuracy and efficiency of individual stages, while the contribution lies in the test-reuse formulation and validation pipeline.

\paragraph{Dependence on AI models.}
\toolname{} uses GPT-4o for mutation candidate generation and assertion adaptation. In contrast, test reproduction, candidate execution, assertion checking, and dynamic validation are performed directly on the running application without AI inference. Thus, the AI model is used to propose candidate behavioral variations and adapt their assertions, while the validity of these variations is determined through actual app execution. Better AI models may improve the quality of candidate generation and assertion adaptation, but the runtime validation mechanism remains independent of the underlying AI model.
\section{Related Works} \label{sec:related}

\textbf{Voice Assistants and Mobile GUI Agents.}
Recent advances in Voice Assistant (VA) systems have focused on automated task completion by interpreting GUI screen contexts and aligning them with user requests~\cite{vu2023voicify}. In HCI, smartphone VAs have been studied for decades, often through Universal Control and Programming by Demonstration (PBD). Universal Control approaches support direct application control at the mobile OS level~\cite{JustSpeak, voiceAccess, VoiceOver} or through intelligent conversations~\cite{vu2023voicify, DoThisHere, bhalerao2017smart, sec2act}. PBD systems~\cite{Sugiura, Cypher} use record-and-replay mechanisms~\cite{VASTA, Kite, PUMICE, SUGILITE, AutoVCI} to let users create shortcuts for routine app features. More recent LLM-based systems extend this line by translating textual prompts into smartphone actions~\cite{huang2025prompt2task}, improving mobile-agent task completion through page-reaching and operation reasoning~\cite{wu2025reachagent}, and using structured UI maps to support app-specific automation~\cite{lin2025uicompass}. Prior work has also explored deriving app-specific conversational capabilities from GUI tests~\cite{qin2025apptova}. \toolname{} builds on this direction by discovering behavioral variations within tests and generalizing validated variations into reusable VA capabilities.

\textbf{Code Extraction and Mutation.}
The closest techniques to our code extraction problem fall under code mining, which has been used for design template mining~\cite{4796208,8115652,6676875}, specification mining~\cite{MiningSpecifications,Pradel_ASE09,Nguyen_FSE09}, and bug detection~\cite{Acharya_FSE07,PR-Miner,MAPO}. These techniques usually require large code corpora and recurring patterns, while GUI test suites often contain only a small number of customized task scripts. Mutation testing offers a more suitable foundation because it systematically creates behavioral variants for validation. Traditional Java mutation tools~\cite{MuJava,Javalanche,PIT} and Android mutation tools~\cite{GUIMutate,Edroid} focus on program faults or GUI component properties. Recent work has begun using LLMs to generate more semantic mutants~\cite{tip2025llmorpheus} and intent-level mutations~\cite{hamidi2025intentmutation}, while Android-focused mutation frameworks such as METFORD address the engineering cost of mobile mutation testing~\cite{vincenzi2025metford}. In contrast, \toolname{} mutates GUI test events to discover valid behavioral variations, rather than faults, and uses dynamic validation to decide which variations can be exposed as VA capability parameters.

\textbf{Test Reuse and Migration.}
Code reuse~\cite{Frakes} repurposes existing software artifacts to reduce development effort. In GUI testing, reuse has often been studied as test migration across platforms~\cite{testmig,Talebipour,Jun-Wei} or across applications~\cite{Behrang,Behrang2,Leonardo}. Recent LLM-based migration approaches transfer Android tests across apps by abstracting source tests into natural-language descriptions~\cite{beyzaei2025llmigrate}, extracting test intentions before migration~\cite{cao2025item}, using VLM-based agents with active feedback~\cite{li2025reusedroid}, or separating GUI test migration into abstraction and concretization stages~\cite{zhang2025macdroid}. These approaches reuse tests by moving them to another app or platform. 
Unlike prior app-specific VA work that maps a GUI test to a fixed capability, \toolname{} identifies invariant and variable interactions within the test and generates a parameterized capability only when the variations are dynamically validated.

\section{Conclusion} \label{sec:conclusion}

In this paper, we proposed \toolname{}, an approach that generalizes existing GUI tests into reusable app-specific VA capabilities through mutation-based exploration and dynamic validation.
Across 396 reproduced events from 100 GUI tests, \toolname{} reduced 18,999 mutation candidates to 1,272 and identified mutable events with an 85.50\% F1 score.
Among 65 generated capabilities, 64 executed successfully and 56 satisfied our quality requirements.
A user study with 18 participants and 150 observations achieved an 82.0\% VA success rate, with 16 participants rating the VA as useful or very useful.
Overall, the results demonstrate the feasibility of generalizing test-derived task logic into reusable app-specific VA capabilities.

\bibliographystyle{ACM-Reference-Format}
\bibliography{sample-base}

\end{document}